\documentclass[russian]{rcdj}

\renewrcdthm[remark]{rem}{Comment}

\renewcommand{\l}{\lambda }
\renewcommand{\t}{\tilde}

\renewcommand{\phi}{\varphi}
\renewcommand{\t}{\tilde}

\begin{document}

\setcounter{page}{449}
\journal{REGULAR AND CHAOTIC DYNAMICS, V.\,8, \No4, 2003}
\runningtitle{MOTION OF A CIRCULAR CYLINDER AND $\bs n$ POINT VORTICES IN A PERFECT FLUID}
\title{MOTION OF A CIRCULAR CYLINDER AND $\bs n$ POINT VORTICES IN A PERFECT FLUID}
\runningauthor{A.\,V.\,BORISOV, I.\,S.\,MAMAEV, S.\,M.\,RAMODANOV}
\authors{A.\,V.\,BORISOV}
{Institute of Computer Science\\
Universitetskaya, 1\\
426034, Izhevsk, Russia\\
E-mail: borisov@ics.org.ru}
\authors{I.\,S.\,MAMAEV}
{Institute of Computer Science\\
Universitetskaya, 1\\
426034, Izhevsk, Russia\\
E-mail: mamaev@ics.org.ru}
\authors{S.\,M.\,RAMODANOV}
{Moscow, Russia\\
E-mail: ramodanov@mail.ru}

\abstract{The paper studies the system of a rigid body interacting dynamically with point vortices in
a perfect fluid. For arbitrary value of vortex strengths and circulation around the cylinder the system
is shown to be Hamiltonian (the corresponding Poisson bracket structure is rather complicated). We
also reduced the number of degrees of freedom of the system by two using the reduction by symmetry
technique and performed a thorough qualitative analysis of the integrable system of a cylinder interacting
with one vortex.}
\amsmsc{76B47, 37J35, 70E40}
\doi{10.1070/RD2003v008n04ABEH000257}
\received 29.10.2003.

\maketitle

In this paper we consider the system of a  rigid cylinder  interacting
with point vortices. We start by indicating some known results on the
subject from the classical hydrodynamics. For the most part, these results
are presented in~\cite{Vill, Lamb, Sefmen}.

As far as we can see, Kirchhgoff~\cite{Kirhgoff} was the first who studied
the dynamics of point vortices  on a systematic basis. In particular, he
obtained the equations of motion in Hamiltonian form and indicated
integrals of motion. It was enough to show that the equations of  motion
governing the system of two or three vortices are integrable (the problem
of three vortices was first analytically solved by Gr\"obli).

The classics also considered  the system of point vortices moving
externally to rigid, stationary boundaries (the impermeable condition on
the boundaries was assumed). Greenhill~\cite{Greenhill} studied in detail
the motion of two vortices in a circular region. Havelock~\cite{Havelock}
investigated stability of n-gon stationary configurations in the region
exterior to a circle. We should also mention F\"oppl's analysis of
interaction of a rigid body with an ambient flow at low Reynolds numbers
(more exactly, he   investigated  stability of the system of two vortices
interacting with a circular cylinder embedded in a uniform flow).

At the same time, the study of the dynamics of  interacting vortices-body
systems has been also a subject of interest in the classical
hydrodynamics. It is known from the phenomenological theory developed by
Prandtl~\cite{Prand} and Jukowski that when a body moves in a fluid, the
thin boundary layer peels off the body thereby generating bound vortices.
In their turn, these vortices exert  a lifting force on the body which can
be observed in aero- and hydrodynamical experiments.

The problem of interaction of  a rigid body and point vortices in a
perfect fluid (the impermeable condition on the body's surface is assumed)
can be studied within the framework of the Hamiltonian mechanics. Various
forms of the equations of motion for a rigid circular cylinder interacting
with $n$ point vortices have been recently (and practically
simultaneously) obtained in ~\cite{ram,ramo,shashi}. The integrability of
the equations in the case of one vortex ($n=1$) was established
in~\cite{BorMam_2}.

In this paper we will study in greater detail  this case of integrability.
We will also consider the simplest chaotic system of a  cylinder and two
vortices to which  a reduction procedure will be applied resulting in
reduction  of degrees of freedom by one.

Hereinafter, as in~\cite{BorMam_2}, the term {\it rigid body} will refer
to  a two-dimensional circular region. It should be noted that even in the
case of an elliptic region the equations of motion become much more
complicated and cannot be written in such a compact form.

\section{Hamiltonian form of the equations of motion}

The equations of motion for a cylinder and vortices with respect to a
fixed coordinate frame $Oxy$ can be written as~\cite{ram} \vspace{-1mm}
\eqc[ur1]{
\left.\dot{\br}_i=-\bv{+}\grad\wt\vfi_i\right|_{\br=\br_i},\q \dot \br_c = \bv\\
a\dot v_1=\lm v_2{-}\suml_{i=1}^n\lm_i(\dot{\wt y}_i{-}\dot y_i),\;
a\dot v_2={-}\lm v_1{+}\suml_{i=1}^n\lm_i(\dot{\wt x}_i{-}\dot x_i),
}
where $\br_c$~ is the radius\1vector from $O$ to the center of mass of the
cylinder, $\bv$~is the velocity of the cylinder, $\br_i$~is the vector
from the center of the cylinder to the $i$\1th vortex and   $\dot{\wt
\br}_i = \frac{R^2}{{\wt \br}_i}\br_i$ is the vector from the center of
the cylinder to  the $i$\1th inverse point (Fig.~1). Here $R$ denotes the
radius of the cylinder, the constant coefficient $a$ involves the added
mass of the cylinder; and the constants $\lm$ and~$\lm_i$ are connected
with the circulation around the cylinder and the vortex strengths  by the
formulae $\lm=\frac{\Gamma}{2\pi}$, $\lm_i=\frac{\Gamma_i}{2\pi}$. The
density of the fluid is $2\pi$.

\fig<bb=0 0 71.0mm 47.6mm>{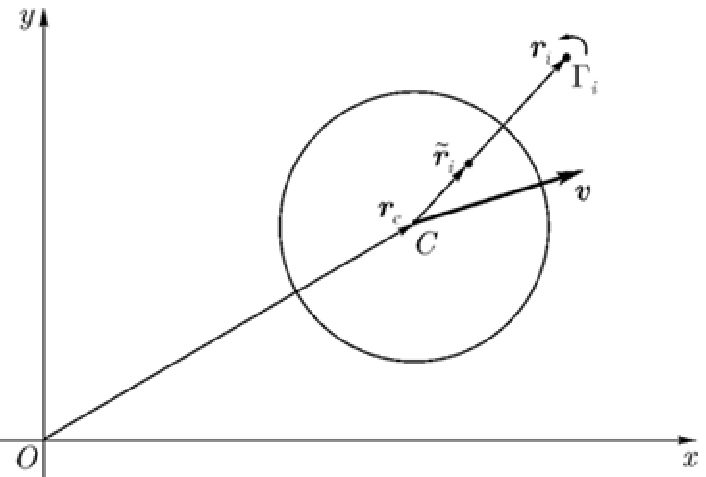}

The function  $\wt {\vfi}_i(\br)$ represents that portion of the velocity
potential ${\vfi}_(\br)$ which does not have a singularity at the
point~$\br = \br_i$. The velocity potential in the region exterior to the
cylinder reads
\eqa{
\label{eq2}
\vfi(\br)=-\frac{R^2}{\br^2}\,(\br,\bv)-\lm\arctg\frac yx
+\suml_{i=1}^n\lm_i\Bigl(\arctg\Bigl(\frac{y-\wt y_i}{x-\wt x_i}\Bigr)-
\arctg\Bigl(\frac{y-y_i}{x-x_i}\Bigr)\Bigr).
}

Equations~\eqref{ur1} were derived using a balance of linear momentum for
the fluid within a circular boundary that encloses the body and  the
vortices. The fluid is assumed to be at rest at
infinity~\cite{ram,shashi}.

Thus the analysis of the system of a cylinder and vortices in a perfect
fluid can be reduced to the analysis of a finite set of ordinary
differential equations. It is easy to check that equations~\eqref{ur1}
preserve invariant measure, i.e., the divergence of~\eqref{ur1} is zero.

As for the system of  $n$ point vortices~\cite{Kirhgoff},
equations~\eqref{ur1} can be shown to be Hamiltonian.
\begin{pro}
Equations~\eqref{ur1} can be represented in the form
\eq[eq3]{
\dot \zeta_i = \{\zeta_i,\,H\} = \suml_{k} \{\zeta_i,\,\zeta_k\}\pt{H}{\zeta_k}
}
where $\zeta_i$ are the components of the phase vector $\bs \zeta =
(x_c^{},\,y_c^{},\,v_1^{},\,v_2^{},\,x_1^{},\,y_1^{}\dts x_n^{},\,y_n^{})$
and  $H$~is the Hamiltonian. For the components $J_{ij}^{}(\bs \zeta) =
\{\zeta_i^{},\,\zeta_j^{}\}$ of the structural tensor of the Poisson
bracket structure the Jacobi identity holds:
\eq*{
\suml_l\left(J_{il}\pt{J_{jk}^{}}{\zeta_l^{}} +
J_{kl}\pt{J_{ij}^{}}{\zeta^l}+
J_{jl}\pt{J_{ki}}{\zeta^l}\right)=0\,\qq \forall i,\,j,\,k
}
\end{pro}

\proof  Equations~\eqref{ur1} has an integral of motion:
\begin{equation}
\label{H}
H=\frac12\,a\bv^2 + \frac12\suml_i
\Bigl(\lm_i^2\ln(\br_i^2-R^2)-\lm_i\lm\ln \br_i^2\Bigr)
+\frac12\suml_{i<j}\lm_i\lm_j\ln
\frac{R^4-2R^2(\br_i,\,\br_j)+\br_i^2\br_j^2}{|\br_i-\br_j|^2}.
\end{equation}
This integral  resembles the Hamiltonian of the system of $n$ point vortices~\cite{Kirhgoff}.

Let  $H$ be our Hamiltonian. Now  we have to  find a skew-symmetric tensor
$J_{ij}^{}$ such that the equations of motion~\eqref{eq3} coincide
with~\eqref{ur1}.  The non-zero components of this tensor are
\eq[ur3]{
\begin{gathered}
\{v_1,\,x_i\}=\frac1a\frac{r_i^4{-}R^2(x_i^2{-}y_i^2)}{r_i^4},\q
\{v_1,\,y_i\}=-\frac1a\frac{2R^2x_iy_i}{r_i^4},\\
\{v_2,\,x_i\}=-\frac1a\frac{2R^2x_iy_i}{r_i^4},\q
\{v_2,\,y_i\}=\frac1a\frac{r_i^4+R^2(x_i^2-y_i^2)}{r_i^4},\\
\{v_1,\,v_2\} = \frac\lm{a^2}{-}
\sum_i\frac{\lm_i}{a^2}\frac{r_i^4-R^4}{r_i^4},\q \{x_i,\,y_i\}=-\frac1{\lm_i},\\
\{x_c,\,v_1\}= \{y_c,\,v_2\} = a^{-1}.
\end{gathered}
}
It is easy to verify that for the Poisson bracket~\eqref{ur3} the Jacobi
identity holds.\qed

The Poisson bracket structure~\eqref{ur3} is non-degenerate, therefore,
by the Darboux theorem, it can be reduced to a canonical form
($\{q_i^{},\,p_j^{}\} =\delta_{ij}$). However, in our further analysis
canonical coordinates  will not be used.

The Lie-Poisson bracket structure for~\eqref{ur1}, \eqref{eq2} under the
condition~$\lm = - \suml \lm_i$  was studied in \cite{shashi}. In this
work stability of an equilibrium configuration of the system of two
vortices behind a steadily moving  cylinder  is discussed. The Poissom
bracket~\eqref{ur3} was first obtained in~\cite{BorMam_2}.

The fact that the general equations are Hamiltonian is not a priori
obvious and does not seem to follow from the Lagrangian formalism. The
Hamiltonian form of the equations allows us to apply the highly developed
perturbation theory techniques (e.g. the KAM theory) and other specific
methods of qualitative analysis. The Liouville theorem on integrability
and its geometrical extension suggested by Arnold~\cite{Arnold} can be
applied to the analysis of these equations.

\section{Problem of advection}

The problem of finding pathlines of the fluid for a given  motion of the
cylinder $\br_c^{} = \br_c(t)$ and the vortices $\br_i^{} = \br_i(t)$,
$i=1\dts n$ is known as {\it the problem of advection}. As mentioned
above, the ambient flow is potential with potential~\eqref{eq2}, therefore
the equation of motion for a passive particle with respect to the
cylinder-fixed frame of reference looks like
\eq[eq6]{
\dot \br = \grad \vfi(\br)\bigr|_{\br_i^{} =
\br_i(t)},\qq \br = (x,\,y).
}

Obviously, equations~\eqref{eq6} are Hamiltonianian with respect to the
standard Poisson bracket structure $\{x;\,y\}=1$; the Hamiltonian is
time-dependent and coincides with the stream function for the
potential~\eqref{eq2}, that is,
\eqa[eq7]{
H_a(\br,\,t) =& \left( \frac{R^2}{r^2}-1\right)\left(v_1^{}(t)y - v_2^{}(t)x\right) +
\frac{1}{2}\lm \ln \br^2 +\\
& + \frac{1}{2} \suml_{i=1}^N\lm_i\left(\ln |\br - \br_i(t)|^2 - \ln |\br -
\wt \br_i(t)|^2\right).
}

\section{Symmetry and integrals of motion}

The equations of motion~\eqref{ur1} are invariant under the action of the
Euclidean group~$E(2)$, therefore, by Noether's theorem for
Hamiltonian systems, there exist three integrals of motion.

The integrals
\eq[eq7.5]{
Q = a v_2^{} + \lm x_c^{} - \suml \lm_i^{}(\wt x_i^{} - x_i^{}),\qq
P = a v_1^{} - \lm y_c^{} + \suml \lm_i^{}(\wt y_i^{} - y_i^{})
}
correspond to  translations along the coordinate axes. These integrals are
a generalization to the classical linear  momentum. On the other hand, the
vector $(Q,\,P)$  can be considered as a counterpart of the vector of {\it
the center of vorticity} of the system of $n$ vortices and cylinder. For
the system of $n$ vortices this notion was introduced in~\cite{Kirhgoff,
Lamb}. For $\lm \ne 0$, the origin of coordinates can be so chosen as to
make  $Q = P = 0$, meaning that the center of vorticity can be always
shifted to  the origin of coordinates.

The third integral, corresponding to the invariance under  rotations about
an axis perpendicular to the plane of motion,  is
\eq[eq8]{
I = a(v_1^{}y_c^{} - v_2^{}x_c^{}) - \frac{1}{2}\lm \br_c^2 -
\frac{1}{2}\suml_{i=1}^n \lm_i^{} \br_i^2 + \frac{1}{2}
\suml_{i=1}^{n}\left(\frac{R^2}{\br_i^2} - 1\right)(\br_i^{},\,\br_c^{}).
}

In the paper~\cite{BorMam_2}, an additional integral for
equations~\eqref{ur1}  but with the dynamics for the center of the
cylinder excluded  was indicated. This integral looks like
\eq[F]{
\begin{gathered}
F=a^2\bv^2+\suml_{i=1}^n\lm_i\biggl(2a\biggl(1{-}\frac{R^2}{\br_i^2}\biggr)
(x_iv_2{-}y_iv_1){+}(\lm_i{-}\lm)\br_i^2{+}\lm_i\frac{R^4}{\br_i^2}\biggr)+\\
+2\suml_{i<j}\lm_i\lm_j(\br_i,\br_j)\biggl(1{-}\frac{R^2}{\br_i^2}\biggr)
\biggl(1-\frac{R^2}{\br_j^2}\biggr).
\end{gathered}
}
The integrals $F$,$I$, $P$ and $Q$ satisfy the equation
\eq*{
F = 2\lm I + P^2 + Q^2 + 2R^2 \suml_{i=1}^{n} \lm_i^2.
}

\begin{rem*}
The Hamiltonian vector field that corresponds to the integral~\eqref{eq8} looks like
\eq[eq8.5]{
\bX_I^{}=\{\bs \zeta,\,I\} = (y_c^{},\,-x_c^{},\,v_2^{},\,-v_1^{},\,y_1^{},\,-x_1^{}\dts
y_n^{},\,-x_n^{})
}
\end{rem*}

The Poisson bracket of the integrals $Q$, $P$, and $I$ differs from the
Lie\1Poisson bracket for the $e(2)$ algebra by a constant
(co-cycle~\cite{Arnold}), that is,
\eq[eq9]{
\{Q,\,P\} = \lm,\qq \{I,\,Q\} = P,\qq \{I,\,P\} = -Q.
}

{\it Thus, for $\lm \ne 0$ the number of degrees of freedom of the system
governed by~\eqref{ur1} can be reduced by two and even by three if $\lm =
0$ and $P = Q = 0$}.

\begin{cor}
The dynamics of a cylinder and one vortex is integrable in the Liouville sense.
\end{cor}

\begin{cor}
For $\lm = 0$ and  $P = Q = 0$, the dynamics of a cylinder and two vortices is integrable in the Liouville sense.
\end{cor}

\section{Complex form of equations of motion and the Dirac bracket}

As mentioned above, in~\cite{BorMam_2} reduced equations were used, i.e.,
the equations we dealt with were exactly equations~\eqref{ur1} but without
the equation $\dot \br_c = \bv$. The elimination of this equation from the
general system~\eqref{ur1} now can be interpreted as the reduction by
symmetry  due to the integrals $Q$ and $P$~\eqref{eq7.5}. For  $\lm \ne 0$
the reduction can be carried out by restricting the dynamics to a joint
level surface of the integrals~\eqref{eq7.5}. Since  $\lm \ne 0$, we
assume that the origin of the fixed coordinate frame  is at the center of
vorticity, i.e., $P=Q=0$. Then,  we substitute     $\dot \br_c$ for
$v_i^{}$ in~\eqref{eq7.5}. The first order equations in the positional
variables result. It is convenient to write these equations in the complex
form
\eq[z]{
\begin{gathered}
a\dot z_c^{} = av =-i\lm z_c+i\sum\limits_{j=1}^n\l_j\left(
\t z_j- z_j\right)\\
\dot{\bar{z}}_k = -\bar{v} + \frac{R^2 v}{z_k^2} +
i\frac{\lm}{z_k^{}}-i\lm_k^{}\frac{1}{z_k^{} - \wt z_k^{}}+
\suml_{j \ne k}^n \lm_i^{}\left(\frac{1}{z_k^{} - z_j^{}} -
\frac{1}{z_k^{}-\wt z_{j}^{}}\right),\\
\wt z_k^{} = \frac{R^2}{\bar{z}_k^{}}\qq k = 1\dts n,
\end{gathered}
}
where  $z_c^{} = x_c^{} + i y_c^{}$ and $z_k^{} = x_k^{} + iy_k^{}$ define
the position of the cylinder's center and the vortices, and $v = v_1^{} +
iv_2^{}$~is the velocity of the cylinder's center.

Obviously, equations~\eqref{z} are Hamiltonian. The Hamiltonian can be
obtained by replacing the cylinder's velocity in~\eqref{H} with the
expression in the right-hand side of the first equation of~\eqref{z}
\eqa[eqq15]{
H = & \frac{1}{2a}\left(\lm x - \suml_{i=1}^n\lm_i(\wt x_i^{} - x_i^{})\right)^2 +
\frac{1}{2a}\left(\lm y - \suml_{i=1}^n\lm_i(\wt {y}_i^{} - y_i^{})\right)^2 +\\
+ & \frac12\suml_i
\Bigl(\lm_i^2\ln(\br_i^2-R^2)-\lm_i\lm\ln \br_i^2\Bigr)\,+
\frac12\suml_{i<j}\lm_i\lm_j\ln
\frac{R^4-2R^2(\br_i,\,\br_j)+\br_i^2\br_j^2}{|\br_i-\br_j|^2}.
}

The Poisson bracket for~\eqref{z} is
\eq[eq11]{
\{x_c^{},\,y_c^{}\} = \frac{1}{\lm}, \qq \{x_i^{},\,y_i^{}\} = -
\frac{1}{\lm_i^{}},\qq i=1\dts n.
}
This bracket can be obtained via the Dirac reduction
procedure~\cite{Dirak}, which consists in restricting the
bracket~\eqref{ur3} to the manifold $Q = P = 0$. In other words, for $\lm
\ne 0$, the cylinder can be considered as an $(n+1)$\1th "compound
vortex", and equations~\eqref{z} govern the system of $n+1$ vortices.

\begin{rem*}
The Dirac bracket~$\{\cdot, \cdot\}_D^{}$ on a manifold ${\cal N}_c$ which
is a  level surface  $f_i(\bs \zeta) = c_i$, $c_i = \const$, $i = 1\dts k$
is given by the formula
\eq*{
\{g,\,h\}_D = \{g,\,h\} + \suml_{ij}\{g,\,f_i^{}\} c_{ij}\{h,\,f_j^{}\},
}
where $\|c_{ij}\| = \|\{f_i,\,f_j\}\|^{-1}$ and $\{\cdot,\,\cdot\}$~is
the original Poisson bracket. In our case we have
\eq*{
\{g,\,h\}_D = \{g,\,h\} - \frac{1}{\lm}\left(\{g,\,Q\}\{h,\,P\}-
\{g,\,P\}\{h,\,Q\}\right),
}
where $\{\cdot,\,\cdot\}$ is the Poisson bracket~\eqref{ur3}
\end{rem*}

Equations~\eqref{z} are invariant under rotations about the origin (the
center of vorticity) and therefore have an additional integral of motion:
\eq[eq12]{
I = \lm \br_c^2 - \suml_{i=1}^n \lm_i^{} \br_i^2.
}

\wfig<bb=0 0 63.9mm 64.7mm>[21]{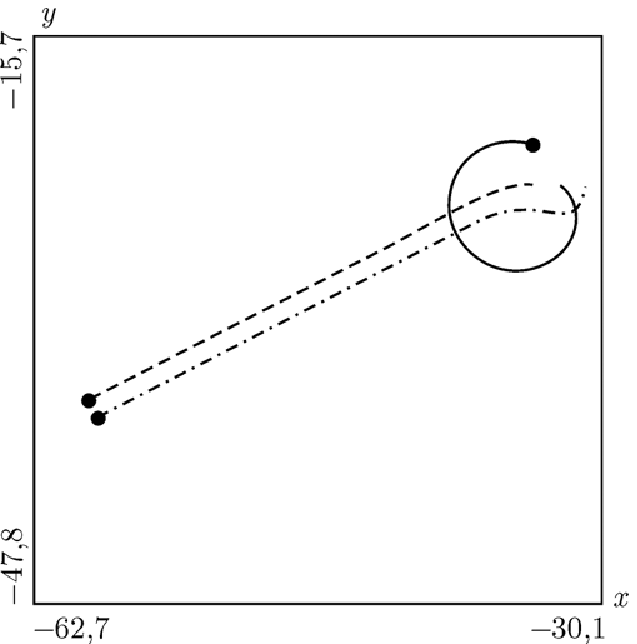}[Motion of a cylinder (solid
line) and a vortex couple ($\Gamma_1=-\Gamma_2$); $a=20$, $\Gamma=8$,
$\Gamma_1=10$, $\Gamma_2=-10$, $R=1$, $F=2$; the initial conditions
(relative to the center of the cylinder) are $v_1(0)=0$, ${v_2(0)=0}$,
$x_1(0)=-3$, $y_1(0)=0$, $x_2(0)=3$, $y_2(0)=0$.]

\noindent
This integral can be obtained  upon substitution of the expression for $\br_c^{}$ from~\eqref{z} for  $\bv$ in the integral~\eqref{F}.

Now we can formulate some properties of the motion of the system of a
cylinder and $n$ vortices. These properties are analogous to those
formulated by Synge~\cite{Synge} for the system   of $n$ vortices.

\begin{pro}
Suppose that the vortex strengths are of the same sign and $\lm\lm_i^{} <
0$; then the trajectory of the cylinder's center and the vortices belong
to a bounded region.
\end{pro}

\begin{pro}
Suppose that  $\lm \ne 0$ and the vortex motions relative to center of the
cylinder are bounded (i.e., $\forall i$ $|\br_i^{}(t)|$ is a bounded
function), then the absolute motion of the cylinder is also bounded.
\end{pro}

Here and in the sequel, a motion with respect to the fixed frame will be
referred to as {\it an absolute motion} and with respect to the center of
the cylinder as {\it a relative motion}.

\begin{rem*}
It should be noted that for $n=1$ and $\l\ne 0$ a more
general statement is valid: the absolute motion of the cylinder is bounded
if and only if the relative motion of the vortex is bounded. Simulations
have shown that already for $n=2$ this is not true: the trajectories of
the cylinder and two vortices for the case $\l_1={-\l_2}$, $|\l|<|\l_1|$ are
shown in Fig.~\ref{unbounde.eps}. As might be expected, eventually the vortices
start drifting to infinity while the cylinder moves along a curve which
gradually takes the shape of a circle.
\end{rem*}

\section{Motion of a cylinder and one vortex}

\paragraph*{Integrability of the equations of motion. Reduction to a system with one degree of freedom.}
Consider in greater detail the system of a cylinder interacting
dynamically with only one vortex, i.e., $n=1$. Let $\br_1=\br=(x,y)$. With
the assistance of the first integrals~\eqref{eq7.5} and~\eqref{eq8}, the
solution to the equations of motion follows in terms of quadratures. Using
the integrals, our system can be reduced to a system with one degree of
freedom. The {\it algebraic} reduction that we will now use is analogous
to the Routh reduction.

As the variables of the reduced system, it is reasonable to choose
integrals of the field of symmetry  ~$\bv_I$~\eqref{eq8.5} (see, for
example,~\cite{bormam}). We put \vspace{-1mm}
\eqa[redvar]{
p_1&=a(xv_1+yv_2),\qq p_2=a(xv_2-yv_1),\qq \rho=x^2+y^2.
}
The Poisson brackets for these variables are
\begin{equation}
\label{ur7}
\begin{gathered}
\{p_1,p_2\}=(\lm-\lm_1)\rho+\frac1{\lm_1}(p_1^2+(p_2-\lm_1R^2)^2),\\
\{p_1,\rho\}=2\rho+\frac2{\lm_1}(p_2-\lm_1R^2),\qq \{p_2,\rho\}=-2\frac{p_1}{\lm_1}.
\end{gathered}
\end{equation}

In terms of \eqref{redvar} the integrals~\eqref{H} and \eqref{F} read
\begin{gather}
\label{H1}
H={p_1^2+p_2^2\over
2a\rho}+\frac{1}{2}\l_1^2\ln\left(\rho-R^2\right)-\frac{1}{2}\l_1\l\ln \rho,\\
\label{F1}
F={p_1^2+p_2^2\over
\rho}+2\l_1\left(1-{R^2\over
\rho}\right)p_2+\l_1^2\left(\rho+{R^4\over \rho}\right)-\l\l_1\rho.
\end{gather}

The rank of the Poisson bracket structure~\eqref{ur7} is two, i.e., the
structure is degenerate. Hence, the reduced system has one degree of
freedom. One variable of the set~$(p_1,\,p_2,\,\rho)$ can be eliminated
using the integral~\eqref{F1}, which is a Casimir function for the
structure~\eqref{ur7}.\looseness=1

Traditionally~\cite{Arnold}, on a two-dimensional level surface of the
Casimir function (a symplectic leaf), canonical coordinates ($\{q,\,p\} =
1$) are introduced. The reduced equations are
\eq*{
\dot{q} = \pt{H}{p},\qq \dot{p} = - \pt{H}{q}.
}

In contrast to this traditional approach, the local coordinates  that we
will use account well for the leaf's geometry but are not  canonical.  The
phase portraits in terms of these coordinates are vivid and illustrative.
Our attempts to find a simple and natural set of canonical coordinates
have not been successful.

\paragraph*{General properties of motion.}

The reduced system~\eqref{ur7} describes the motion of the vortices relative to the center of the cylinder. Before proceeding to a qualitative analysis of the reduced system, we will indicate some general properties of the absolute
\mbox{motion}.\looseness=1

\begin{pro}
The absolute motion of the cylinder is bounded except maybe for the two
cases:\smallskip

$1.$ $\lm=\lm_1$;\smallskip

$2.$ $\l=0$;
\end{pro}

\begin{rem*}
Condition (1) corresponds to the case where the circulation around the
cylinder is zero,  and condition (2) represents the case where the
circulation around the cylinder and the vortex is zero.
\end{rem*}

\proof
Assume the contrary:  $\l\ne\l_1$, $\l\ne 0$ and the motion of the cylinder is not bounded. Then, according to Prepositions  2 and 3,
$\rho$ is an unbounded function of time and $\l\l_1>0$.
Dividing \eqref{F1} by $\rho$ yields
\eq[sheet]{\left(\l_1+\frac{p_2}{\rho}-\frac{\l_1R^2}{\rho}\right)^2
+\frac{p_1^2}{\rho^2}-\left(\l\l_1-\frac{2\l_1^2R^2-F}{\rho}\right)=0}

Obviously,
\eq*{
\lim\limits_{\rho\to\infty}\left [\left(\l_1+\frac{p_2}{\rho}\right)^2
+ \left(\frac{p_1}{\rho}\right)^2\right ]=\l\l_1.
}
Therefore, there exist functions $\alpha(\rho)=o(1)$ and $\beta(\rho)$
such that $\l_1+\frac{p_2}{\rho}=\sqrt{\l_1\l+\alpha}\cos\beta,\; \frac{p_1}{\rho}=\sqrt{\l_1\l+\alpha}\sin\beta. $

Upon substitution of these expressions into~\eqref{H1}, we get\vspace{-1mm}
\eq*{
2H=\frac{\rho}{2a}k+\l_1\ln\frac{(\rho-R^2)^{\l_1}}{\rho^\l}
}
The second term in the right-hand side is of  $O(\ln \rho)$, the factor
$k$ is separated from zero, that is,
$k=(\l_1-\sqrt{\l_1\l})^2+\alpha+2\l_1(\sqrt{\l_1\l}-
\sqrt{\l_1\l+\alpha}\cos\beta)>k_1=\const>0$. This is a contradiction with
the fact that $H$ is constatnt. \qed

\begin{rem*}
One can easily note that for $\lm = \lm_1$ and $\lm = 0$ an unbounded
motion of the cylinder always exists. Therefore the above preposition
serves as a criterion  of boundedness of the absolute motion. Most likely
this criterion remains valid for the case of an arbitrary (finite) number
of vortices, but this is not proved yet.
\end{rem*}

The statement given below provides a nice "geometro\1dynamical"\;  insight
into the structure of  absolute motions of the cylinder.

\begin{teo*}
Suppose that $\l\ne 0$; then for each periodic solution of the reduced system  \eqref{redvar} there exists a rotating coordinate frame with the origin at the center of vorticity such that with respect to this frame  the vortices and the cylinder move along closed curves.
\end{teo*}

\proof It follows from~\eqref{z} that
\eq[eq18]{yx_1-xy_1=\frac{p_1}{\l},\quad yy_1+xx_1=-\frac{p_2}{\l}
+\frac{\l_1}{\l}(R^2-\rho).}
 By the assumption of the theorem, the functions in the right-hand side of these equations are periodic functions with period $T$.
 Therefore\vspace{-1mm}
\eq[link]{
\left(\begin{array}{c} x_1\\ y_1\end{array}\right)=
{\bs \Phi}\cdot\left(\begin{array}{c} x\\ y\end{array}\right),\quad
{\bs \Phi}=\left(\begin{array}{cc} \Phi_1 & \Phi_2\\-\Phi_2 & \Phi_1 \end{array}\right)
\vspace{-1mm}
}
where $\Phi_1$ and $\Phi_2$ are periodic functions with period $T$.
Substituting these expressions into  \eqref{z}, we get\vspace{-1mm}
\eq[period] {
a\dot x=\l
y-yG_2+xG_1,\quad a\dot y =-\l x +yG_1+xG_2,
\vspace{-1mm}
}
where $G_1$ and $G_2$ are also  $T$\1periodic functions.
Let ${\bf A}$ be the matrix for the system of linear differential equations
 \eqref{period}. It can be easily verified that
${\bf A}\cdot\int\limits_0^t{\bf A}\,dt= \int\limits_0^t {\bf A}\,dt\cdot {\bf A}$,
hence  (see, for example,~\cite{Demidovich})
the fundamental matrix, ${\bf X}$, of equations \eqref{period} can be written
as\looseness=-1
\vspace*{-1mm}
$$
{\bf X}=e^{\int\limits_0^t{\bf A}\,dt}.
\vspace{-1mm}
$$
Let us represent ${\bf A}$  as a sum\vspace{-1mm}
$$
{\bf A}={\bf B}+{\bf C},
\vspace{-1mm}
$$
\vspace{-1mm}
$$
{\bf B}=\frac{1}{a}\left( \begin{array}{cc} 0 &
\l-\<G_2\>  \\
 -\l+\<G_2\>  & 0
\end{array}\right),\quad
{\bf C}=\frac{1}{a}\left( \begin{array}{cc}
G_1 & -G_2+\<G_2\>  \\
 G_2-\<G_2\>  & G_1
\end{array}\right).
\vspace{-1mm}
$$
Note that $\<G_1\>=0$, otherwise equations~\eqref{period} have unbounded
solutions which is in a contradiction with Preposition~3. Since the
matrices ${\bf B}$ and  ${\bf C}$ commute, we have\vspace{-1mm}
$$
{\bf X}=\left( \begin{array}{cc} \cos \frac{\l-\<G_2\>}{a}t & \sin
\frac{\l-\<G_2\>}{a}t  \\
 -\sin\frac{\l-\<G_2\>}{a}t & \cos \frac{\l-\<G_2\>}{a}t
\end{array}\right)\cdot {\bf G.}
\vspace{-1mm}
$$
Here ${\bf G}$ is a $T$\1periodic matrix. Thus, with respect to a
coordinate frame (with origin at the center of vorticity) rotating at a
rate $(\l-\<G_2\>)/a$ the cylinder moves along a closed curve. Let us
prove now that the vortex also moves along a closed curve. Let $x^*,y^*$
be the coordinates of the vortex in the fixed coordinate frame. Then
$x^*=x_c^{}+x,\: y^*=y_c^{}+y $. It follows from~\eqref{link}
that\vspace{-1mm}
\eq*{
\begin{gathered}
\left(\begin{array}{c} x^*\\ y^*\end{array}\right)=
({\bs \Phi}+{\bf E})\cdot\left(\begin{array}{c} x\\ y\end{array}\right)=\\
= \left( \begin{array}{cc} \cos \frac{\l-\<G_2\>}{a}t & \sin
\frac{\l-\<G_2\>}{a}t  \\
 -\sin\frac{\l-\<G_2\>}{a}t & \cos \frac{\l-\<G_2\>}{a}t
\end{array}\right)\cdot ({\bs \Phi}+{\bf E})\cdot {\bf G}\cdot
\left(\begin{array}{c} x(0)\\ y(0)\end{array}\right)
\end{gathered}
\vspace{-1mm}
}
Here ${\bf E}$ is the identity matrix, and the matrix  $({\bs \Phi}+{\bf
E}) {\bf G}$ is  $T$\1periodic. \qed

\fig<bb=0 0 117.3mm 53.5mm>{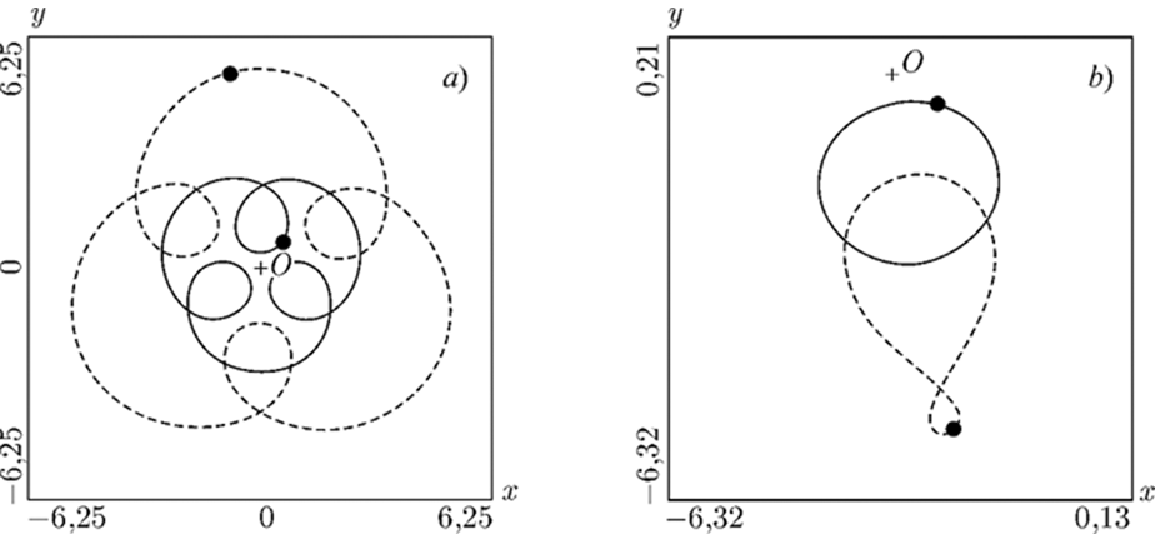}[{\it a}) Trajectories of
the cylinder (solid line) and the vortex (here the two frequencies of the  motion are rationally related).
$a=6,\,\Gamma=-3,\,\Gamma_1=1,\,R=1,\,F=2,\,$;
the initial conditions (relative to the center of the cylinder) are
$v_1(0)=0.24,\,v_2(0)=0,\,x(0)=0,\,y(0)=1.31$;
{\it b}) The trajectories shown in diagram
{\it a} in the frame of reference rotating at a
rate 0.029 (the origin, $O_+$, is at
the center of vorticity).]

An absolute motion of the cylinder (solid line) and the vortex (dashed
line) is shown in Fig.~\ref{chor_e.eps}\,{\it a}.  In
Fig.~\ref{chor_e.eps}\,{\it b} this motion is shown in a rotating frame of
reference whose origin, $O_+$, is at the center of vorticity. The value of
the physical parameters for this motion are
$a=6,\,\Gamma=-3,\Gamma_1=1,\,R=1$, $F=2$ and the initial conditions are
$v_1(0)=0.24,\,v_2(0)=0,\,x(0)=0,\,y(0)=1.31$.

\paragraph*{Qualitative analysis of the reduced system.}
The geometry of the symplectic leaf~\eqref{F1} of the Poisson bracket
structure \eqref{ur7} (the phase space of the reduced system) is governed
by $\l\l_1$. The symplectic leaf is compact if $\l\l_1<0$ and non-compact
if either $\l\l_1>0$ or $\l=0$. Let us consider these three cases.

{\bf 1. Compact case ($\l\l_1<0$).} From \eqref{F1} it follows that
\eq[compact]{\left(\l_1\rho+p_2-\l_1R^2\right)^2+p_1^2+\left(\sqrt{-\l\l_1}
\rho+K\right)^2=K^2,\quad K=\frac{2\l_1^2R^2-F }{
2\sqrt{-\l\l_1}}
}
The symplectic leaf is diffeomorphic to a two-dimensional sphere. For
real motions ${F\ge2\l_1R^2(\l_1-\l)}   $. Since the phase space is compact, ~$\rho$ is a bounded function, meaning that the distance between the vortex and the cylinder cannot grow infinitely. Moreover, in view of Preposition 3, the absolute motion of the
cylinder also is bounded.

On the leaf we intoduce coordinates $\phi$ and $\psi$ by the formulae\nopagebreak\vspace{1.5mm}
\eq[param]{
\begin{gathered}
\l_1\rho+p_2-\l_1R^2=K\cos\phi\cos\psi,\quad p_1=K\sin\phi\cos\psi,\\
\rho\sqrt{-\l\l_1}+K=K\sin\psi,\quad \psi\in\left(-\frac{\pi}{2},\frac{\pi}{2}\right),\; \phi\in\left[-\pi,\pi\right].
\end{gathered}
\vspace{1.5mm}
}

\begin{rem*}
For a fixed value of $K$, the coordinate  $\psi$ satisfies the relation ${\rho {=} \frac{K(\sin \psi -1)}
{\sqrt{-\lm \lm_1}} > R^2}$.
\end{rem*}

To find stationary solutions of the reduced system, consider the differential equations in  $p_1,p_2$
and $\rho$:\vspace{1.5mm}
\eq[redeq]{
\begin{gathered}
\dot p_1=\{p_1,H\}=\frac{ \l_1 R^4\l a\rho-2R^2\l_1\l
a\rho^2-\l_1^2a\rho^3+\l_1^2R^2a\rho^2+\l_1\l a\rho^3}{
\rho^2a(-\rho+R^2)}\\[2pt]
 +\frac{(-R^4\l_1\rho+\l
a\rho^2+R^6\l_1-\l_1R^2\rho^2-\l R^2a\rho+\l
R^2\rho^2-\l_1a\rho^2+\l_1\rho^3-\l \rho^3)p_2}{\rho^2a(-\rho+R^2)}\\[2pt]
+\frac{(\rho^2-R^4)p_2^2+(\rho^2-2R^2\rho+R^4)p_1^2}{\rho^2a(-\rho+R^2)},\\[2pt]
 \dot p_2{=}\{p_2,H\}{=}\frac{(\l
\rho^3{-}\l_1\rho^3{+}\l R^2a\rho{-}R^6\l_1{+}\l_1a\rho^2{-}\l a\rho^2{-}\l
R^2\rho^2{+}\l_1R^2\rho^2{+}R^4\l_1\rho)p_1} { \rho^2a(-\rho+R^2) }+ \\[2pt]
 +\frac{(-2R^2\rho+2R^4)p_2p_1}{\rho^2a(-\rho+R^2)},\\[2pt]
 \dot
\rho=\{\rho,H\}=\frac{(-2\rho+2R^2)p_1}{ a\rho}
\end{gathered}
\vspace{1.5mm}
}
The last equation implies $p_1=0$ and the right-hand side of the second
equation becomes zero. Therefore, on the phase portrait, all stationary
solutions belong to the lines $\psi=\pm\frac{\pi}{2},\,\phi=0$ and
$\phi=\pm\pi$. To determine $p_2$ and $\rho$ we should use~\eqref{compact}
and equate to zero the right-hand side of the first equation. The number
of stationary solutions depends on~$F$.\looseness=-1

Equations~\eqref{redeq} remain unaltered under the transformation
$(p_1,\,p_2,\,\rho,\,t,\,\l,\l_1)\to
(-p_1,-\,p_2$, $\rho,\,-t,-\,\l,-\l_1)$. In view of~\eqref{param}, this
transformation implies the following change of variables and physical
parameters $(\phi,\,\psi,\l,\l_1)\to (\phi+\pi,\,\psi,-\l,-\l_1)$.
Therefore, a change in sign of the circulation and the vortex strength
results in a shift of the phase curves along the $\phi$\1axis by $\pi$.
Thus, with no loss of generality, we can assume that $\l<0$ and $\l_1>0$.

Phase portraits for various values of  $F$ ( the value of the other
parameters are fixed  $a=20$, ${\Gamma=-1},\,\Gamma_1=0.5$ and  $R=1$) are shown
in Figs.~\ref{compcase_e.eps}\,{\it a},~\ref{compcase_e.eps}\,{\it b}
and~\ref{compcase_e.eps}\,{\it c}.
The ``non-physical'' area ($\rho\le R^2$) is shown in grey.

\fig{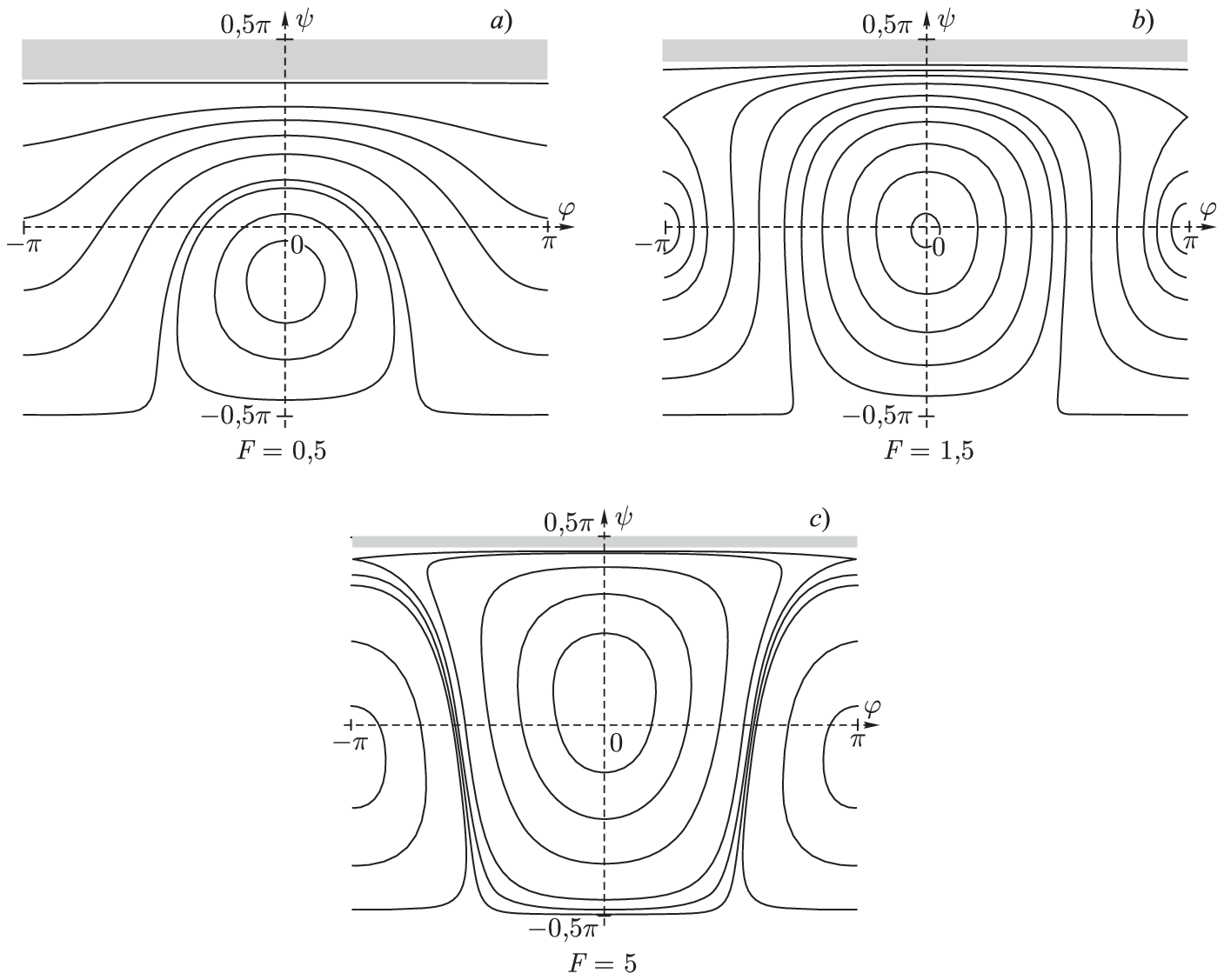}[Phase
portraits of the reduced system in the compact case ($\lm\lm_1<0$):
$a=20$, $\Gamma=-1$, $\Gamma_1=0,5$, $R=1$.]

The solution curves of the reduced equations are the level curves
of~\eqref{H1} on the surface~\eqref{F1}.  The level curves are closed,
hence the solutions of the reduced equations are periodic.

The fixed points on the phase portraits (stationary solutions) represent a
motion in which  the cylinder and the vortex move along concentric circles
whose centers are at the center of vorticity. For example, the
trajectories of the vortex and the cylinder that correspond to the
elliptic point in Fig.~\ref{compcase_e.eps} are shown in
Fig.~\ref{asynhron.eps}. With an increase in $F$ two more fixed points
appear, the motion corresponding
to these points is qualitatively shown in~Fig.~\ref{synchron.eps}.

\ffig<bb=0 0 28.7mm 28.9mm>{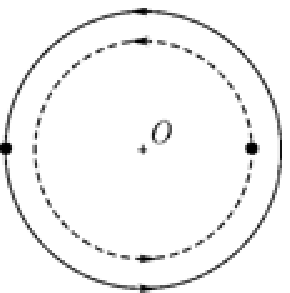}<bb=0 0 28.5mm 28.3mm>{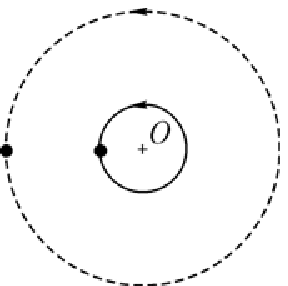}

{\bf 2. Non-compact case ($\l\l_1>0$).} It follows from ~\eqref{sheet}
that
\eq[noncompact]{\left(\l_1\rho+p_2-\l_1R^2\right)^2+p_1^2-\left(\sqrt{\l\l_1}
\rho-K\right)^2=-K^2,\quad K=\frac{2\l_1^2R^2-F }{
2\sqrt{\l\l_1}}.
}
Symplectic leaves are diffeomorphic to a hyperboloid of two sheets. For
real motions  ${F{>}2\l_1R^2(\l_1{-}\l)}$. On symplectic leaves we introduce
local coordinates:\nopagebreak\vspace{-2mm}
\eq[param1]{
\begin{gathered}
\l_1\rho+p_2-\l_1R^2=K\cosh\phi\sinh\psi,\quad p_1=K\sinh\phi,\\
\rho\sqrt{\l\l_1}-K=-K\cosh\psi\cosh\phi,
\end{gathered}
\vspace{-2mm}
}
Arguing as in the compact case, we can assume that  $\l>0$ and $\l_1>0$.
In view of \eqref{redeq} and \eqref{param1}, on the phase portrait, all
fixed points lie on the axis $\phi=0$. It is interesting to note that,
unlike the compact case, the topology of the phase portrait is determined
by the sign of the difference $\lm-\lm_1$ and  does not change
qualitatively as the constant $F$ varies. Typical phase portraits
are given in Fig.~\ref{noncompc_e.eps}\,{\it a}
($\l=\l_1$),~\ref{noncompc_e.eps}\,{\it b} ($\l<\l_1$)
and~\ref{noncompc_e.eps}\,{\it c}~(${\l>\l_1}$).

\fig{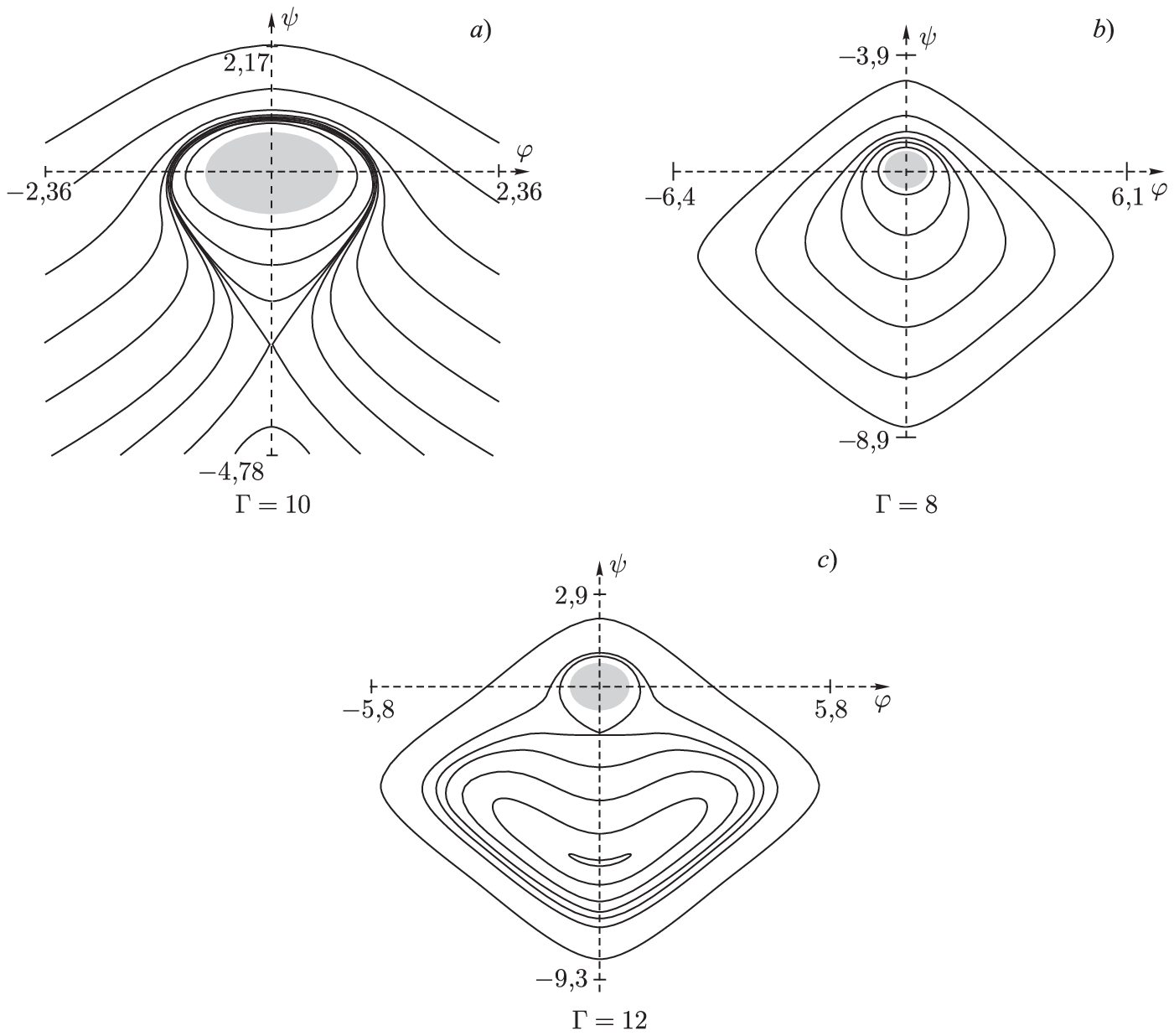}[Phase portraits of the reduced system in the
non-compact case ($\lm\lm_1>0$):
$F=25$, $a=20$, $\Gamma_1=10$, $R=1$.]

The phase curves are closed, hence the solutions of the reduced equations
are periodic functions of time. The only exception is the case $\l=\l_1$
(Fig.~\ref{noncompc_e.eps}). As shown above, only in this case the cylinder
may have an unbounded motion (Preposition 4).

{\bf 3. Non-compact case ($\l=0$).} In the system of equations~\eqref{ur1}, the equations governing the motion of the vortex are uncoupled from the first  equation. Indeed, it follows from~\eqref{eq7.5}
that\vspace{1.5mm}
$$
av_1=-\l_1y_1f(x_1,y_1)+c_1,\quad
av_2 =\l_1x_1f(x_1,y_1)+c_2,
\vspace{1.5mm}
$$
where $f(x_1,y_1)=-1+R^2/(x_1^2+y_1^2)$ and $c_1,\,c_2$ are arbitrary constants.
Substituting these expressions into the first equation~\eqref{ur1}, we obtain a closed system of equations in the unknowns $x_1$ and $y_1$. The solution curves of this system coincide with  level curves of the integral~\eqref{H1}. It can be shown that the level curves are closed, hence $x_1$ and $y_1$ are periodic functions of time. The evolution of the center of the cylinder is governed by the
equations\vspace{1.5mm}
$$
\begin{gathered}
x=\frac{1}{a}\left(-\l_1\left\<y_1f(x_1,y_1)\right\>+c_1\right)t+g_1(t),
\\[6pt]
y=\frac{1}{a}\left(\l_1\left\<x_1f(x_1,y_1)\right\>+c_2\right)t+g_2(t),
\end{gathered}
$$
where $g_1(t),\, g_2(t)$ are periodic functions. Thus, there exists  a uniformly moving coordinate system in which the orbits of the cylinder and the vortex are closed.

\section{The case of two vortices}

Suppose now that $n=2$, i.e., we are
going to consider the system of a cylinder interacting dynamically with
two vortices. By analogy with the case $n=1$, we use the
integral~\eqref{F} to reduce the number of degrees of freedom by one. As
before, we take some quantities invariant with respect to rotations about
the cylinder's center as the new variables, namely,\vspace{-2mm}
$$
\begin{gathered}
p_1=a(x_1v_1+y_1v_2),\quad p_2=a(x_1v_2-y_1v_1),\quad p_3=x_1x_2+y_1y_2,
\quad p_4=x_1y_2-x_2y_1,\\
r_1=x_1^2+y_1^2,\quad r_2=x_2^2+y_2^2.
\end{gathered}
$$
The Poisson brackets for these variables are as follows:
\eqa*{
\{p_1,p_2\}& =\frac {(  p_2^2 +   p_1^2) r_2^2 - 2\l_1R^2  p_2 r_2^2
+ (R^4 -  r_1^2)\l_1^2 r_2^2 + r_1^2\l\l_1 r_2^2 +
r_1^2\l_1\l_2(R^4 -  r_2^2)}{\l_1 r_1 r_2^2},\\
\{p_1,p_3\}& =\frac{  r_2^2(  p_3  p_2 -   p_4  p_1) + \l_1( - (R^2
 -  r_1) r_2^2  p_3 + 2R^2
  p_4^2 r_1 - R^2 r_1^2 r_2
 +  r_1^2 r_2^2)}{\l_1 r_1
 r_2^2},\\[-2pt]
\{p_1,p_4\}& =\frac  {r_2^2(  p_1
  p_3 +   p_2  p_4) - \l_1  p_4
(2R^2  p_3 r_1 + (R^2 -  r_1)
 r_2^2)}{\l_1 r_1 r_2^2},\\[-2pt]
\{p_1,r_1\}& =\frac {2  p_2 - 2(R^2 -
r_1)\l_1}{\l_1},\quad
\{p_1,r_2\}=- \frac {2  p_3(R^2 -  r_2
) }{r_2},\\[-2pt]
\{p_2,p_3\}& =\frac { -  r_2^2(  p_1
  p_3 +   p_2  p_4) - \l_1  p_4
(2R^2  p_3 r_1 - (R^2 +  r_1)
 r_2^2)}{\l_1 r_1 r_2^2},\\[-2pt]
\{p_2,p_4\}& =\frac { r_2^2(  p_3
  p_2 -   p_4  p_1) + \l_1( - (R^2
 +  r_1) r_2^2  p_3 - 2R^2
  p_4^2 r_1 + R^2 r_1^2 r_2
 +  r_1^2 r_2^2)}{\l_1 r_1
 r_2^2},\\[-2pt]
\{p_2,r_1\}& = - \frac{ 2  p_1}{\l_1},\quad
\{p_2,r_2\}=- \frac {2  p_4(R^2 -  r_2)}{ r_2},\quad
\{p_3,p_4\}=\frac{ \l_2 r_2 - \l_1
 r_1}{\l_2\l_1},\\[-2pt]
\{p_3,r_1\}& =\frac{ 2  p_4}{\l_1},\quad
\{p_3,r_2\}=- \frac {2  p_4}{\l_2},\\[-2pt]
\{p_4,r_1\}& = - \frac{ 2  p_3}{\l_1},\quad
\{p_4,r_2\}=\frac {2  p_3}{\l_2},\quad
\{r_1,r_2\}=0.
}

\begin{figure}[!ht]
\centering
\cfig{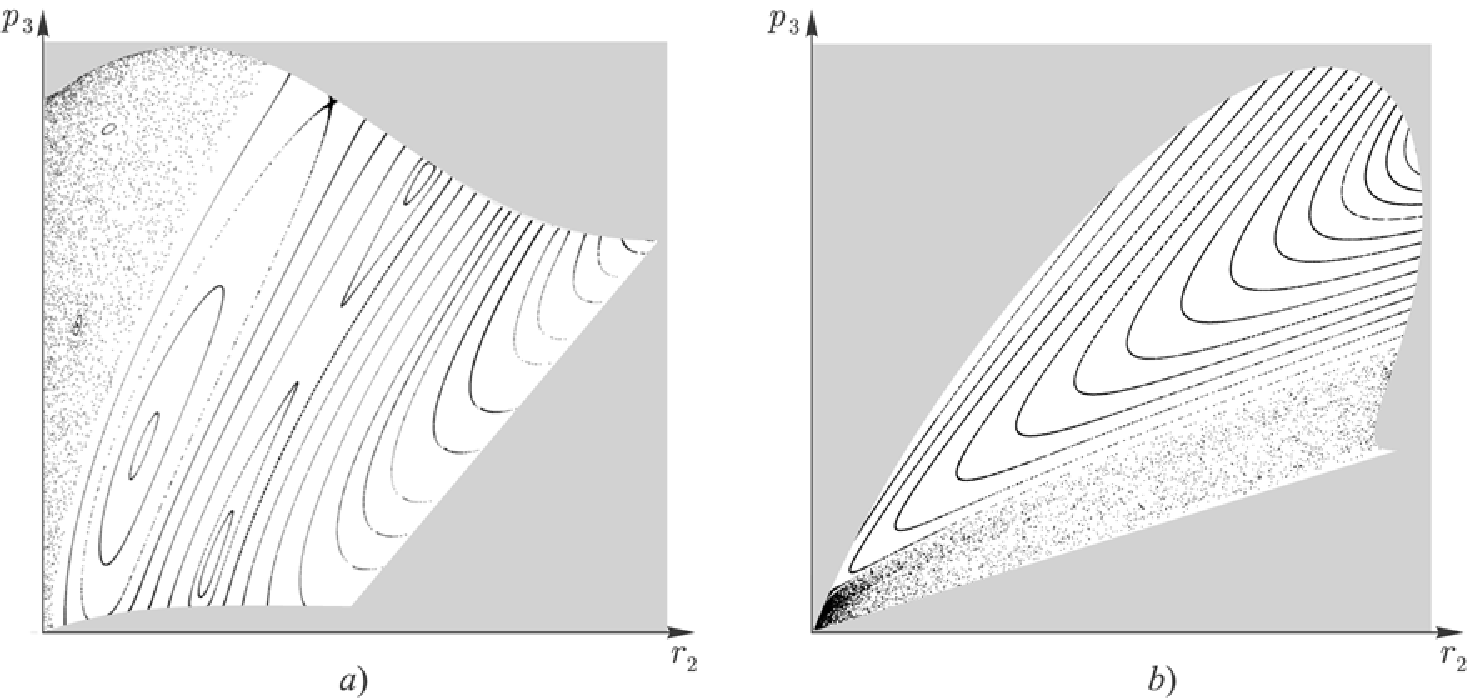}

\caption{Poincar\'e surface-of-section plots for the system of a cylinder and two vortices:
{\it a}) $H=7$, $F=5$, $a=9$, $\Gamma=1$, $\Gamma_1=\Gamma_2=8$, $R=1$, $p_4=-5$;
{\it b}) $H=10$, $F=60$, $a=4$, $\Gamma=2$, $\Gamma_1=-1$, $\Gamma_2=10$, $R=1$,
$p_4=0$.}\vspace{-2mm}
\end{figure}

This Poisson bracket structure is degenerate, and its rank is four. The
integral \eqref{F} and the obvious relation $p_3^2+p_4^2=r_1r_2$ are
Casimir functions for this structure. For the reduced system to be
integrable, one more first integral is needed.

To explore the reduced system numerically, we have used the Poincar\'e
surface-of-section technique. The variables~$r_2,\,p_3,\,p_4$ can be
considered local coordinates on a three-dimensional manifold on which the
integrals~\eqref{H} and \eqref{F} are constant and the equation
$p_3^2+p_4^2=r_1r_2$ is fulfilled; $p_4$ is the cross variable. Two
Poincar\'e surface-of-section plots are given in Fig.~\ref{chaos_e.eps}.
The~chaotic behavior of solutions proves that in the general case an
additional first integral does not exist.\vspace{-2mm}

\section{Conclusion}\vspace{-2mm}

Very often equations of motion that have not been derived within the framework of Lagrangian formalism (i.e., using the calculus of variations) are not Hamiltonian (even though the energy may be a conserved quantity for these equations).
For example, for equations of the nonholonomic mechanics there are dynamical obstacles preventing the existence of a Poisson bracket structure~\cite{BorMam_3}. Since the equations of motion~\eqref{ur1} are Hamiltonian, they have the generic features of Hamiltonian dynamical systems:
for any value of parameters there are no attractors (e.g., strange attractors) in the phase  space; at the same time, there are invariant KAM tori separated with stochastic layers. In this paper, the chaotic system of a cylinder and two vortices has been touched briefly. It seems that it would be interesting to explore some particular motions of this system (both regular and chaotic) in greater detail.

Mention should be made of the paper~\cite{Ramod_2} where a modification of the famous Bjerknes problem, the system of two dynamically interacting 2D cylinders, is considered. The equations of motion for this system are not integrable.

This work was supported in part by
 Leading Scientific School of Russia Support grant
 no.~\mbox{\selectlanguage{russian} อุ-136.2003.1}.

\end{document}